\documentclass{article}

\usepackage{graphicx}
\newcommand{\kk}[2]{\frac{#1}{#2}}
\newcommand{\pab}[2]{\frac{\p #1}{\p #2}}

\def\be{\begin{equation}}
\def\ee{\end{equation}}

\def\p{\partial}

\def\bu{{\bf u}}

\def\Am{\stackrel{\circ}{A}}
\def\Km{W/mK }
\def\={\approx}
\def\8{{\infty}}

\begin{document}

\title{\large \bf Modelling Heat Transfer of Carbon Nanotubes}

\author{Xin-She Yang \\
Department of Engineering, University of Cambridge \\
Trumpington Street, Cambridge CB2 1PZ, UK   }

\date{ }

\maketitle

\begin{abstract}

Modelling heat transfer of carbon nanotubes is important for the
thermal management of nanotube-based composites and nanoelectronic
device. By using a finite element method for three-dimensional
anisotropic heat transfer, we have simulated the heat conduction and
temperature variations of a single nanotube, a nanotube array and a
part of nanotube-based composite surface with heat generation. The
thermal conductivity used is obtained from the upscaled value from
the molecular simulations or experiments. Simulations show that
nanotube arrays have unique cooling characteristics due to its
anisotropic thermal conductivity. \\ \\


\noindent {\bf Citation detail:} Yang, X.-S., 
``Modelling Heat Transfer of Carbon Nanotubes",
{\it Modelling and Simulation in
Materials Science and Engineering}, Vol. {\bf 13}, 893-902 (2005).  

\end{abstract}

\newpage

\section{Introduction}

Carbon nanotubes (CNT) have attracted worldwide attention in both
research and industry since the discovery by Iijima in 1991
\cite{Iijima91}. Extensive studies of this novel nanocomposite
material suggests that it has many potential applications such as
nanomachines, nanoelectronic and nanomedicine due to its super-high
stiffness, strength and resilience, and its exceptional electrical
and thermal properties
(\cite{Iijima93,Cornwell,Dressel,Govin,Goze}). It bas become one of
the hottest research topics in the important nanotechnology. There
have been extensive studies about the properties of carbon nanotubes
(\cite{Arroyo,Belyt,Hali,Treacy,Wong,Yakobson}). In addition, there
have important development concerning the multi-scale computer
simulation techniques to imbue the continuum-based models with more
realistic details at quantum and atomistic scales \cite{Curtin}.
There are many technical challenges such as the experimental studies
of the mechanical, thermal and electronic properties of
CNTs(e.g.,\cite{Demcz,Ruoff,Treacy}). Even though there have been
extensive theoretical modelling studies in term of atomistic,
molecular dynamics (MD), quantum mechanic and continuum approach,
many computational issues remain unsolved or unsatisfied. Among
these important issues are the modelling of the nano-rheological
behaviour and heat transfer of carbon nanotubes, the fluid flow
inside a nanotube or interconnected nanotube networks.

The modelling of nanothermal process such as heat transfer and
thermal stress of carbon nanotubes has interesting challenges. The
complexity of the nanotubes makes the analytical approach
intractable, while the experimental measurement is very expensive
and time-consuming. This makes the computational modelling a better
quick and yet efficient alternative for studies and analysis of the
structure of carbon nanotubes and nanocomposites
(\cite{Das,Li1,Liu,Odegard,Popov,Van,Pipes}). Finite element
analysis has been quite successful in the modelling of macroscale
phenomena such as heat transfer, engineering mechanics and fluid
flow. However, special care should be taken before it can apply to
model the nanoscale structure. In recent years, there have been
several interesting studies using the continuum mechanical method to
study the microscale structure and mechanical properties
(e.g.,\cite{Arroyo,Das,Li1,Liu,Odegard,Prylu}). However, there has
been little progress in the modelling of heat transfer and thermal
management of carbon nanotubes or nanotube-based composites.

This paper aims at the development of the finite element analysis
of the nanoscale heat transfer of carbon nanotubes. We first
formulate the finite element method for 3-D heat transfer analayis
of nanotubes. Then, we study the heat conduction and temperature
variations of single-walled nanotubes, nanotube array and
nanotube-based composites to simulate the thermal behavoir such as
nano-electronic devices and nanoswitch, followed by the discussion
and implication of the simulation results.

\section{Finite Element Formulation}

In the modelling of carbon nanotubes, molecular modelling is a
discrete approach and the conventional finite element method is a
continuum-based approach. Finite element analysis (FEA) on
nanoscales is different from both of these approaches. In fact, it
shall be the combination of these two approaches and bridging
their gaps. Finite element method uses the most elaborate finite
element method while its input materials properties are estimated
from the molecular-based modelling.

\subsection{Scaling Gaps}

Carbon nanotubes and nanocomposites have a length scale range from
the molecular scale to macroscopic scales. There are two
conventional computational methods working on different scales.
Molecular-based computational modelling tries to predict the
mechanical and thermal properties using quantum mechanics and
discrete {\it ab initio} calculations
(\cite{Belyt,Sanchez,Che,Mar,Pipes}), while macroscopic solid
mechanics treats the media as a continuum using homogenized bulk
material properties (\cite{Arroyo,Govin,Zien}). However, the
intermediate nanoscale is just between these two extreme scales. A
proper methodology is needed to bridge these two scales in the
hierarchy.

Such important intermediate scaling of material properties only
start to emerge recently. Odegard et al \cite{Odegard} proposed an
instructive and interesting approach. Another promising method is
molecular mechanical modelling approach (\cite{Weiner},
\cite{Li1}). These methods often use the molecular level phonon
theories, and the thermal conductivity can be derived \cite{Che}.

\subsection{Thermal Conductivity}

The thermal conductivity (K) for carbon nanotubes varies greatly and
it is highly anisotropic as the conductivity is much higher along
the axis of the nanotube than in other directions. Although it is
extremely difficult in obtaining the experimental measurements,
there have some important progress in the experimental studies of
the thermal properties of carbon nanotubes
(\cite{Hone,Small,Ruoff}). Small et al \cite{Small} discussed in
detail the mesoscopic experimental measurements of phonon thermal
transport and thermoelectronic phenomena in individual carbon
nanotubes. They measured the temperature distributions  in
electrically heated individual multiwalled carbon nanotubes with a
scanning thermal microscope and a microfabricated suspended device
to obtain the thermal conductivity.

The bulk thermal conductivity is estimated to be in the range of
$150-6000$ \Km, and it varies with temperature and size of the
nanotubes (\cite{Small},\cite{Che},\cite{Mar}). However, there is a
significant gap between experiment-derived values and theoretical
predictions. Choi et al used 2000 \Km for thermal conductivity
\cite{Choi}, and Kim et al obtained a value of 3000 \Km for the room
temperature thermal conductivity of individual multiwalled nanotubes
\cite{Kim}. Biencuk et al demonstrated that samples loaded with 1
wt$\%$ unpurified single-walled carbon nanotubes showed a 70$\%$
increase in thermal conductivity at 40 K and 125$\%$ at room
temperature \cite{Bier}. Small et al obtained that the 'bulk'
thermal conductivity at room temperature is over 3000 \Km, while the
earlier results suggested a lower value of 36 \Km for the
densely-packed single-walled carbon nanotube mat, and with the
estimated bulk longitude thermal conductivity of a nanorope in the
range of 1600-6000 \Km (\cite{Small,Hone}). In addition, there is
also a large difference between single tube and bulk measurements
and there is more strong temperature dependence in individual
nanotubes than in the bulk measurements as clearly pointed by Small
et al \cite{Small}. Che et al theoretically predicted the thermal
conductivity along the tube axis approaches 2980 \Km for (10,10)
single-walled nanotubes for tube length up to 40nm with a thickness
of 1 $\Am$. They also predicted that a value of 950 \Km for nanotube
bundles along the tube axis and the much lower thermal conductivity
of 5.6 \Km in the direction perpendicular to the tube, and this
value is comparable to the graphite out-of-plane thermal
conductivity 5.5 \Km \cite{Che}. Similarly, the mechanical
properties are also the unusually unique with very high Young's
modulus 1000-3700 GPa and Poisson's ratio 0.18-0.3
(\cite{Kelly,Wong,Demcz,Treacy,Hali,Van}). In fact, the thermal
conductivity of nanotubes is highly anisotropic with the value along
the tube axis usually two orders higher than the perpendicular to
the tube. Thus, the aligned nanotubes, nanoropes and bundles are
also highly anisotropic.

The highly anisotropic properties in thermal conductivity and
large difference between single-tube and bulk measurements suggest
that more extensive studies are needed. From the point of view of
finite element analysis, the formulation shall be specially taken
care of so that the method shall be able to analyze the
anisotropic characteristics in heat transfer of nanotubes. For the
continuum-based finite element method, there have been many ways
to ensure this capability. However, for the finite element
analysis on the nanoscale, some modifications and special choice
of element types should be investigated.  In the rest of the
section, we will formulate the essential implementation of the
finite element analysis for the modelling heat transfer of carbon
nanotubes.

\subsection{Continuum-Based Finite Element Formulation}

Once the parameters of thermal properties are estimated or upscaled
from the molecular simulations or ab initio simulations, an
equivalent continuum-based finite element formulation can be used.
The main procedure is the same as the conventional finite element
procedure, and many finite element algorithms and element types can
be similarly constructed. However, the continuum-based finite
element method only works from the scale of micron to very large
scales \cite{Zien}. For smaller scales, the continuum-based
representative volume does not work on the atomistic scale. There
are two ways to overcome this difficulty. One way is to use the
discrete molecular-mechanical simulations based on the quantum
mechanics, and this may leads to very large numbers of computations
(\cite{Das,Li1,Pipes2}). The other is to use the discrete-based
modelling to upscale the continuum-equivalent material properties
such as thermal conductivity. Then, the normal procedure of finite
element analysis can be used.

For finite element analysis of nanoscale heat transfer of carbon
nanotubes, two natural models for element types  are: truss model
and hexagon model. The truss model is very popular in solid
mechanics for the analysis of engineering structures. Odegard et
al uses a truss model to simulate the mechanical behavior of the
nano-structured systems in terms of the displacement of the atoms
and the total molecular potential energy \cite{Odegard}. The
hexagon model will be studied in more detail in this paper. The
beam and truss models are based on 1-D line model, and thus it is
unlikely to give accurate results for 3-D structures, especially
the geometry is more complex. The anisotropic feature of the heat
conduction of carbon nanotubes requires the full 3-D formulation
of heat transfer equation, and proper initial and boundary
conditions shall be implemented.

In the rest of the paper, we will formulate a full 3-D finite
element method and use the hexagon element with six nodes being
the atoms of the basic carbon structure. We will then simulate the
heat transfer of carbon nanotubes and compare different method for
the same configuration so as to find a better and suitable method
for the finite element method.

The governing equation of 3-D heat conduction can be written as \be
\rho  c_p \pab{T}{t}=\pab{}{x}(K_x \pab{T}{x}) + \pab{}{y}(K_y
\pab{T}{y}) + \pab{}{z}(K_z \pab{T}{z})+Q f(t,T), \ee where $K_x,
K_y, K_z$ are the principal thermal conductivities. $\rho$ and $c_p$
are density and specific heat capacity, respectively. $Q$ is the
rate of heat generation, and $f(t,T)$ is the source term which is a
known function of time $t$ and temperature $T$.

For an ambient temperature $T_{\8}$ and the typical temperature
$T_*$ of the system, any temperature can be written as
$T=T_{\8}+\Theta \Delta T$. The temperature variation is very small
$\Delta T=T_*-T_{\8} \ll T_{\8}$, and typically $\Delta T$ is the
order of 1 K. Equivalently, we can define a dimensionless
temperature \be \Theta=\kk{T-T_{\8}}{T_*-T_{\8}}. \ee For the
ambient room temperature $T_{\8}=300K$ and the typical temperature
$T_*=T_{\8}+1=301K$ of interest, we have $\Theta=1$ when $T=301K$,
and $\Theta=0$ when $T=300K$. In fact, we can choose any temperature
$T_*$ of our interest so that the {\it dimensionless} temperature
difference is $O(1)$.

By choosing the typical length $L$, thermal conductivity $K_0=2500$
\Km, and time scale $\tau \sim L^2 \rho c_p/K_0$, the governing
equation in the dimensionless form becomes \be
\pab{\Theta}{t}=\nabla \cdot (\kappa \nabla \Theta)+\lambda
f(t,\Theta), \ee where $\lambda=Q L^2/K_0 (T_*-T_{\8})$ and
$\kappa=K_{ij}/K_0$ is a $3 \times 3$ matrix depending on the
anisotropic thermal conductivity.  If there is no heat generation,
then $Q=0$ or $\lambda=0$. For a uniform time-independent heat
source, $f(t,\Theta)=1$.

Substituting $\Theta=\sum_{j=1}^N u_j(t) N_j(x,y)$ into the above
equation and after some elaborate calculations, the finite element
formulation usually leads to a generic form \be {\bf M \dot{u} + K
u=F, } \ee and \be \bu=[u_1 \; v_1 \;w_1 \;u_2\; v_2\;
w_2\;...\;u_n\;v_n\;w_n]^T. \ee where ${\bf M}=\int_{\Omega} N_i
N_j d\Omega, \;{\bf K}$ are the coefficient matrix and the
stiffness matrix, respectively. $ F_i=\int_{\Omega} \lambda N_i f
d\Omega +\int_{\partial \Omega} q d\Gamma$ is the contribution of
the heat source and boundary conditions.

\section{Simulations and Results}

Using the finite element method formulated in the above section, we
can simulate some typical characteristics concerning the heat
transfer of carbon nanotubes. We first study the heat conduction of
a single nanotube,  and then we investigate the heat conduction of
an array of parallel nanotubes and the temperature variations of a
nanodevice with heat generation.

Carbon nanotubes have three major types: armchair tubes [e.g.,
(n,n)], chiral tubes [e.g., (8,2)], and zigzag tubes such as (n,0)
(\cite{Dressel},\cite{Van}). We shall focus on one type of the
nanotube such as zigzag. For other types such as chiral or armchair,
the methodology used is the same. Once the detail configuration of
the nanotubes is known, the finite element nodes and elements are
essentially the same. In all the following simulations, we assume
all the carbon bond are in equal length $1.42 \Am$ with equal angles
$120^{\circ}$. The bulk thermal conductivity $K_0=2500 $\Km and the
ratio of transverse to longitude thermal conductivity $\gamma=0.01$.
Thus, the thermal conductivity matrix $\kappa={\rm
diag}(1,1,\gamma)$. The radius $R$ of the carbon nanotube is 1-1.5nm
(normalized in the dimensional form as 1), while the length is 20nm
(normalized as 1), and the thickness of the single-walled carbon
nanotube is taken to 0.34nm (\cite{Popov}, \cite{Van}).

\begin{figure}
\centerline{\includegraphics[height=1.5in,width=2in]{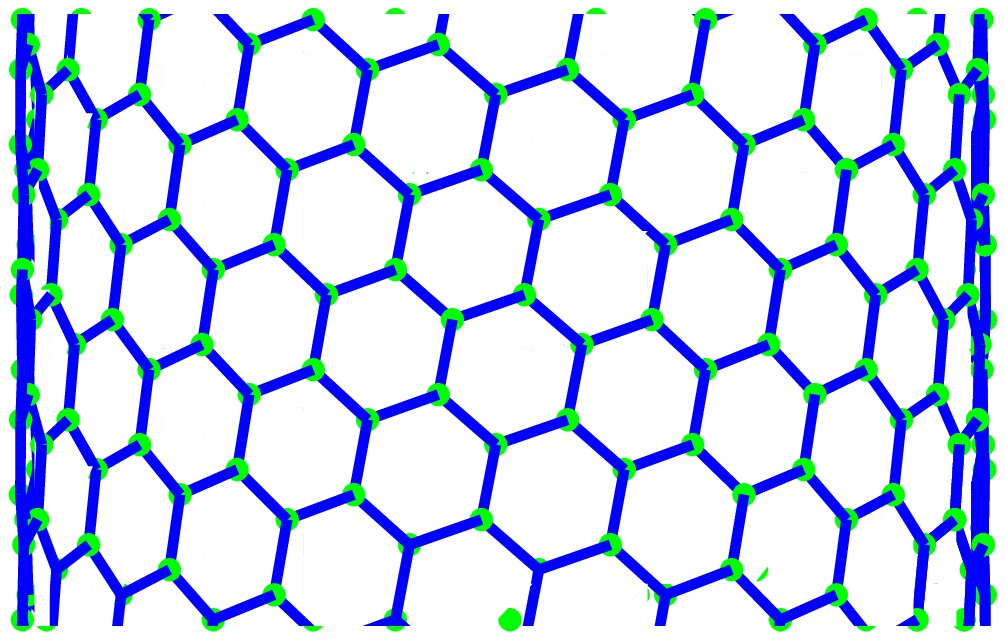}
\includegraphics[height=1.5in,width=1.6in]{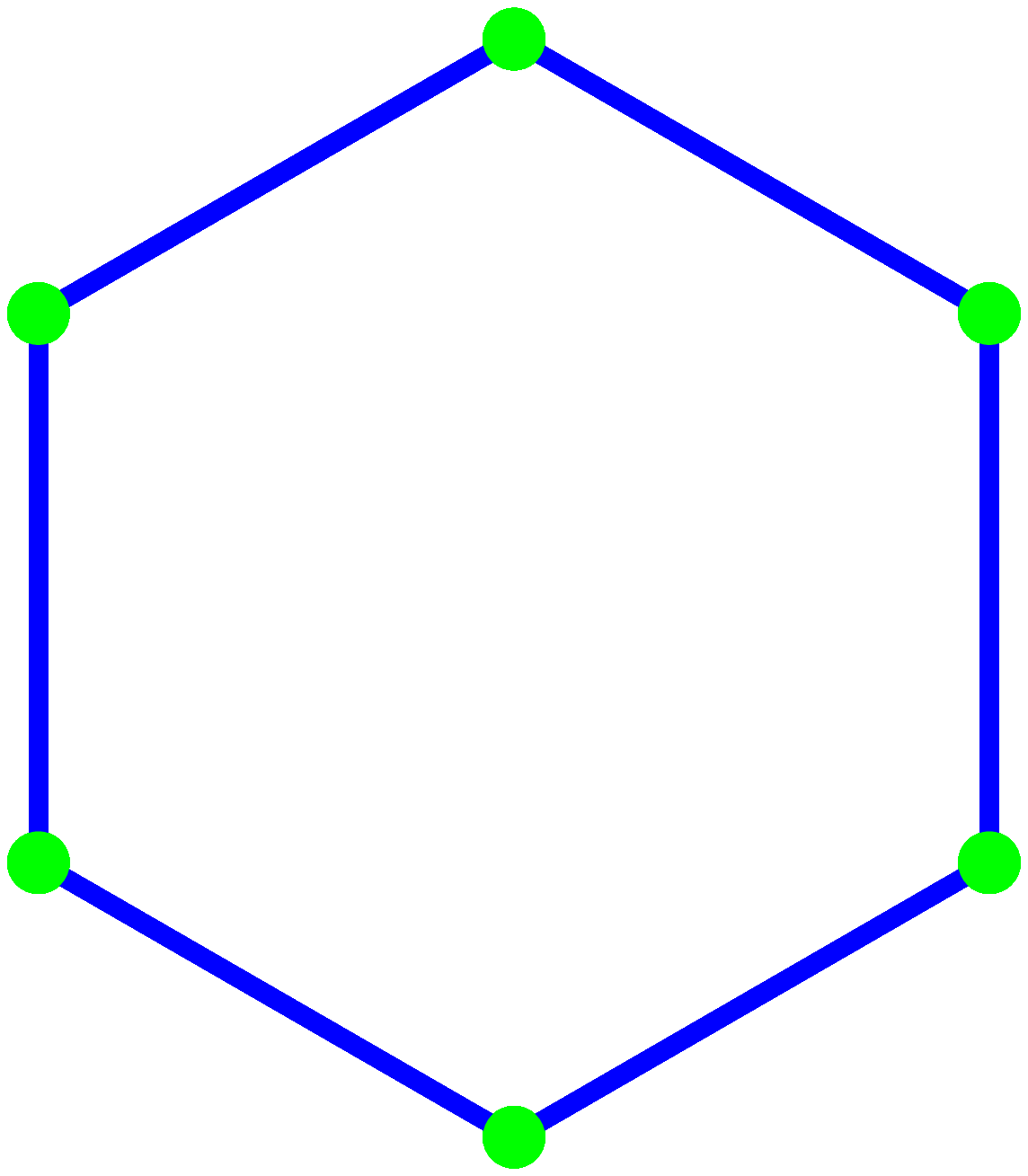}} \caption{A
carbon nanotube and its basic hexagonal structure. }
\end{figure}

In the finite element simulations, we use two major element types:
hexagonal element and beam element. The 6-noded hexagon element is
the same structure as the basic carbon structure (see Figure 1).
We shall present the results using these two element types and
comparison will be made whenever relevant.

\subsection{Heat Conduction of a Nanotube }

Nanoelectromechnical device such as nanoswitches may have many
potential application in nanomedicine and nanoelectronics
(\cite{Deq},\cite{Small}). The thermal behavior of nanodevice and
its heat management is of great interest and practical applications.
For a single nanotube with a radius of 1nm and a length of 20 nm, by
fixing the temperature $\Theta=1$ or $T_0=301K$ at one end and a
radiation boundary at the other end with the ambient temperature
$T_{\8}=300K$, then heat conduction of the nanotube is computed. The
left figure in Figure 2 shows the dimensionless temperature
distribution at time $t=10$. For the same carbon nanotube, if we fix
the temperature at one contact line as $\Theta=1$, then transverse
heat conduction leads to the temperature distribution shown on the
right in the Figure 2. We can see that temperature decrease more
quickly along the tube axis than that in the transverse direction,
and this is because the anisotropic thermal conductivity of the
nanotube.

For the practical purpose of modelling carbon nanotubes, the
hexagonal elements and truss element essentially give essentially
the same results.  Thus, the choice of element types is for the
convenience of computation and programming. However, when the
structure is not thin, the full 3-D finite element method shall be
used, and the hexagonal element generally works better.

\begin{figure}
\centerline{\includegraphics[height=3in,width=5in]{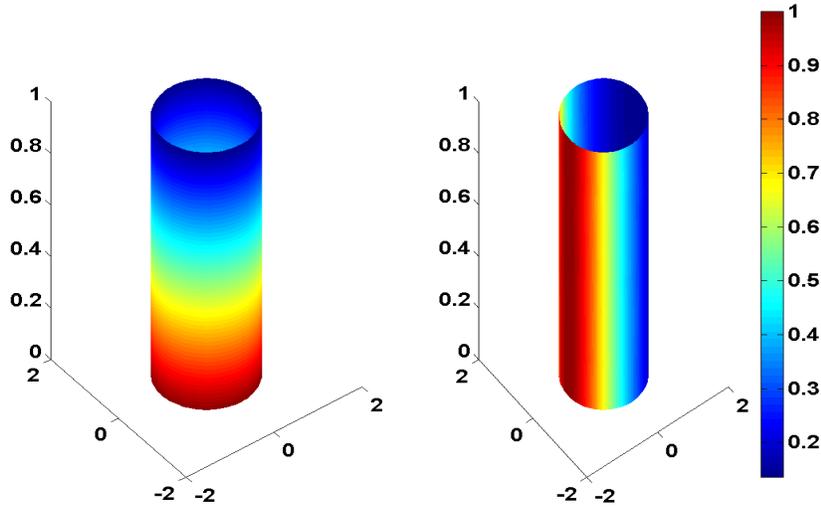} }
\caption{The heat conduction of a carbon nanotube and the
temperature distribution at $t=10$. }
\end{figure}

\subsection{Nanotube Array}

The simulation of an array of parallel carbon nanotubes is of more
practical interest due to the experimental development of carbon
nanotube frabrication and the amorphous carbon deposition
\cite{Gruj}. This makes it more relevant to the nanoscale device.
The heat transfer process on this scale is also important. For an
array of parallelled nanotubes with a radius of 1nm and a spacing of
2.6nm, the base stays at a fixed temperature. The heat conduction of
nanotube array is shown Figure 3 where the fixed base temperature
$\Theta=1$ or $T_0=301K$, and the ambient room temperature
$T_{\8}=300K$ are used.  The normalized temperature distribution is
shown in color calculated using the hexagon elements. Temperature
decrease more quickly along the direction of the axis of the
nanotubes while the transverse temperature variation along the
amorphous base is relatively small. This suggests that an array of
carbon nanotubes can serve as a good cooling device or the nanotube
array-based nanodevice may have very good heat dissipation
properties.

\begin{figure}
\centerline{\includegraphics[height=3in,width=5in]{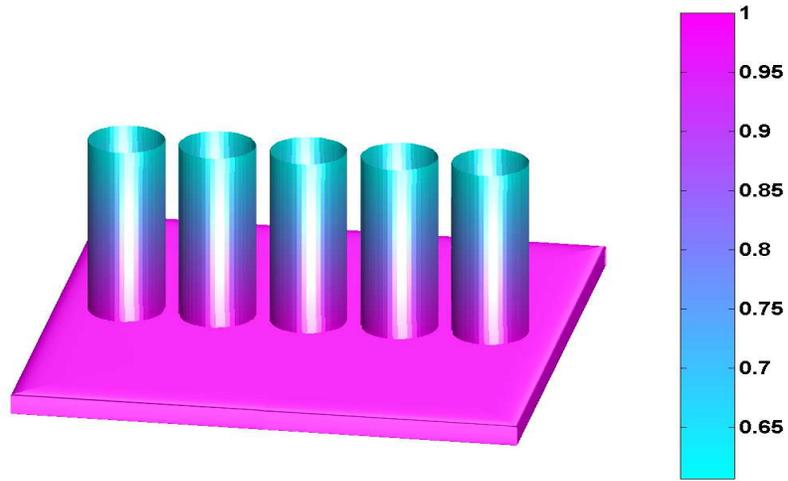} }
\caption{Temperature distribution of a nanotube array at $t=10$. The
temperature decrease more quickly along the direction of the axis of
the nanotubes. }
\end{figure}

\subsection{Nanotube-Based Composite Surface With Heat Generation}

Nanotube-based composites are very promising due to its very unique
mechanical, thermal and nanoelectronic properties. In practice, most
nanoelectronic device will meet the problem of heat management due
to the energy dissipation, heat generation and thermal transport.
For a representative volume of such composite, the continuum-based
simulation can focus on a regular array of nanotubes in parallel to
each other. For the carbon nanotube array discussed earlier, we
embedded them onto a composite surface and add the heat generation
in the system. Figure 4 shows the two cross sections of three
dimensional finite element results for a representative volume with
three carbon nanotubes embedded onto a composite surface. The
composite matrix is assumed to be isotropic and its thermal
conductivity is taken as $500$ \Km. The initial temperature is
$T_0=300.5K$ with a uniform heat generation rate $\lambda=0.2$. The
ambient temperature is $T_{\8}=300K$.

The simulation of this configuration is different from that in
Figure 3 in that the present simulation uses the continuum-based
tetrahedrons and thus the temperature distribution of the nanotubes
is smoothly averaged, while the distribution in Figure 3 is based on
hexagonal elements with the nodes being the same as the atoms. Due
to the large number of atoms involved in the composites, the
hexagonal elements are no longer practical and very time-consuming.
The best choice for composites is the tetrahedron elements. We can
see from Figure 4 that the temperature variations are much bigger
along the nanotubes than in the transverse direction. This is
because the thermal conductivity is much higher along the tube axis
than in the other directions. This interesting characteristic of
heat transfer may imply that the proper aligned nanotubes can have
better cooling mechanism and thus better heat management properties.
This may also imply that these composites can be better choice for
nanoelectronic circuits and other nanomechanical device. From the
above simulations, we see that carbon nanotubes have anisotropic
heat conduction properties. The thermal conductivity is quite high
compared with the usual materials.

\begin{figure}
\centerline{\includegraphics[height=3.5in,width=5in]{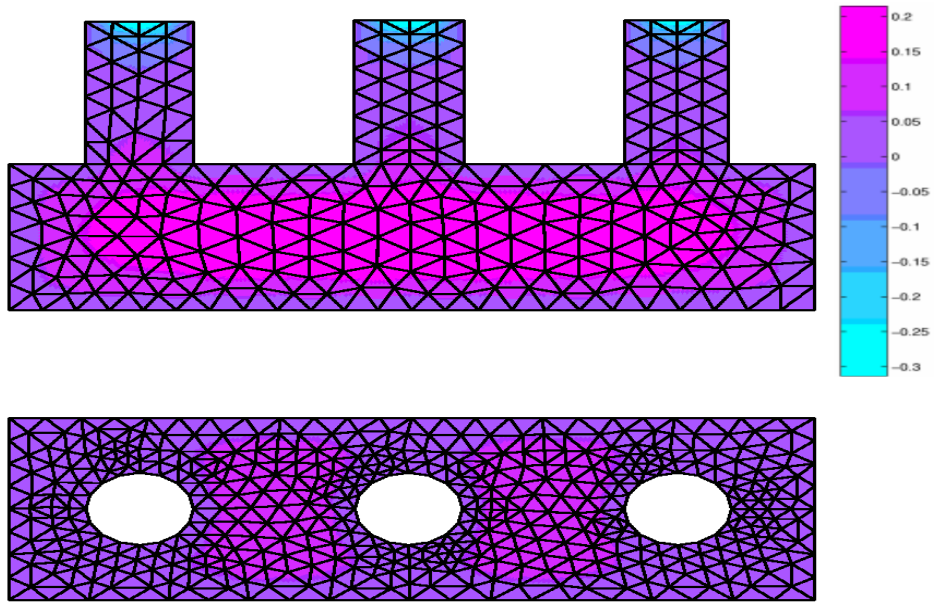} }
\caption{Temperature variation relative to the initial temperature
$T_0=300.5K$ at $t=10$. Heat is generated at a constant rate
$\lambda=0.2$ and the temperature variations are bigger along the
nanotubes than in the transverse direction.}
\end{figure}

\section{Conclusions}

We have formulated a finite element method to simulate the nanoscale
heat transfer of a carbon nanotube and nanotube array. This method
is based on the 3-D continuum-mechanical approach by using material
parameters upscaled from the molecular simulations. The conventional
finite element method usually works from the micron-scale up to a
very large scale. As it reduces to atomistic scales, continuum
approach needs special modifications. These include the upscaling
and continuum-equivalent homogenization of material properties such
as thermal conductivity. For a carbon nanotube with a radius of
1-1.5nm and a thermal conductivity of 2500 \Km and ratio of 0.01
(perpendicular thermal conductivity to axial conductivity), the heat
conduction and temperature variations with heat generation have been
investigated under the possible conditions such as nanomechanical
device. From the natural structure of the carbon nanotubes, we have
used two major element types: the hexagonal elements and beam
elements in the simulations. The comparison of the hexagonal
elements and beam elements shows that for thin structures they give
essentially the same results.

In our simulations, we assume the nanotubes are regular cylindrical
shapes with a uniform thickness. In real nanotubes, caps are an
important part of a carbon nanotube, thus the effect of the cap may
not be negligible. In principle, we can use the the present method
to simulate its effect. In addition, the multi-walled nanotubes and
nanotube-based composites need more extensive studies concerning its
thermal behavior and how their nanothermal behaviours differ from
the single-walled nanotubes. These will become the computational
modelling issues for further research.

{\bf Acknowledgement: } The author thanks the two anonymous referees
for their helpful comments that have improved the manuscript
greatly.

\end{document}